\documentclass[
reprint,
longbibliography,
aps,
twoside,
superscriptaddress,
bibnotes,
floatfix
]{revtex4-2}

\usepackage{physics}
\usepackage{dsfont}
\usepackage{xcolor,newfloat}
\usepackage{graphicx}
\usepackage{dcolumn}
\usepackage{bm}
\usepackage{amsmath,amssymb,amsthm,amsfonts}
\usepackage{mathtools}
\usepackage{multirow}
\usepackage{hyperref}
\usepackage[all]{hypcap}
\usepackage{chngcntr}
\usepackage{enumerate}
\usepackage{nicefrac, xfrac}

\begin{document}

\title{Boosted quantum teleportation}

\author{Simone E. D'Aurelio}
\thanks{These authors contributed equally to this work.}

\author{Matthias J. Bayerbach}
\thanks{These authors contributed equally to this work.}

\author{Stefanie Barz}

\affiliation{Institute for Functional Matter and Quantum Technologies, University of Stuttgart, 70569 Stuttgart, Germany}
\affiliation{Center for Integrated Quantum Science and Technology (IQST), University of Stuttgart, 70569 Stuttgart, Germany.}

\begin{abstract}
Quantum teleportation has proven to be fundamental for many quantum information and communication processes. The core concept can be exploited in many tasks, from the transmission of quantum states, quantum repeaters, to quantum computing.
However, for linear-optical systems, the efficiency of teleportation is directly linked to the success probability of the involved Bell-state measurement. In most implementations, this is realized by linear optics with an intrinsically limited success probability of $50\%$.
Here, we demonstrate quantum teleportation surpassing this limit. We achieve an average fidelity of the teleported states of $0.8677\pm0.0024$, leading to an overall acceptance rate of the teleportation of $69.71\pm0.75\%$.
We obtain this boosted success probability by generating ancillary photonic states that are interfered with the Bell states.
Thus, our work demonstrates the boosting Bell-state measurements in quantum-technology applications and our scheme could directly be applied to e.g. quantum repeaters.
\end{abstract}

\maketitle

\section{Introduction}
Quantum teleportation is an essential building block for quantum communication and for building quantum repeaters~\cite{Bennet1993}. For example,the problem of long-range quantum-state transmission can be reduced
to the distribution of entanglement among participants and the subsequent state teleportation~\cite{Briegel1998}.
Research on quantum teleportation has progressed from proof-of-principle demonstrations~\cite{bouwmeester1997, Popescu1998} to real-world implementations~\cite{ma2012, valivarthi2016, ren2017, llewellyn2020, hermans2022, hu2023} and complex quantum states: teleportation with continuous variables~\cite{xia2018, Zhao2023}, with multiple degrees of freedom~\cite{wang2015}, and of high-dimensional systems~\cite{Luo2019, Hu2020} have been realized.
Teleportation-like protocols can also be used  to implement dense coding~\cite{Mattle96}.\\  
While communication is a natural application of quantum teleportation, the use of quantum teleportation extends beyond this usecase. Measurement-based quantum computation, for instance, relies on teleportation-like operations to perform computation~\cite{daiss2021}, propagation of information, and the generation of cluster states~\cite{bartolucci2023}. The realisation of efficient teleportation operations have therefore the potential to enhance the performance of these systems~\cite{fujii2015,lee2015}.

Photonic systems are a promising technology for performing quantum information tasks, both in communication and in computation. Linear-optical implementations are advantageous due to their experimental simplicity and robustness. However, for quantum teleportation, there exists a fundamental limit in terms of measurement efficiency, as performing a Bell-state measurement (BSM) only succeeds 50\% of all cases~\cite{Luetkenhaus1999, calsamiglia2001,lee2019,choi2020,hilaire2023}. Photonic BSMs can be enhanced in several ways, exploiting for example non-linear effects or hyperentanglement~\cite{Kim2001, Schuck2006, Barbieri2007, Li2017}. A promising alternative is to use linear optics with the addition of ancillary states, which can improve the efficiency of the BSM arbitrarily close to unity by increasing the ancillary-state complexity~\cite{grice2011, ewert2014}. While recent work has demonstrated these advantages experimentally, their performance has not been studied in the context of quantum-information protocols~\cite{bayerbach2023}.
\begin{figure*}
    \centering
    \includegraphics[width=0.8\linewidth]{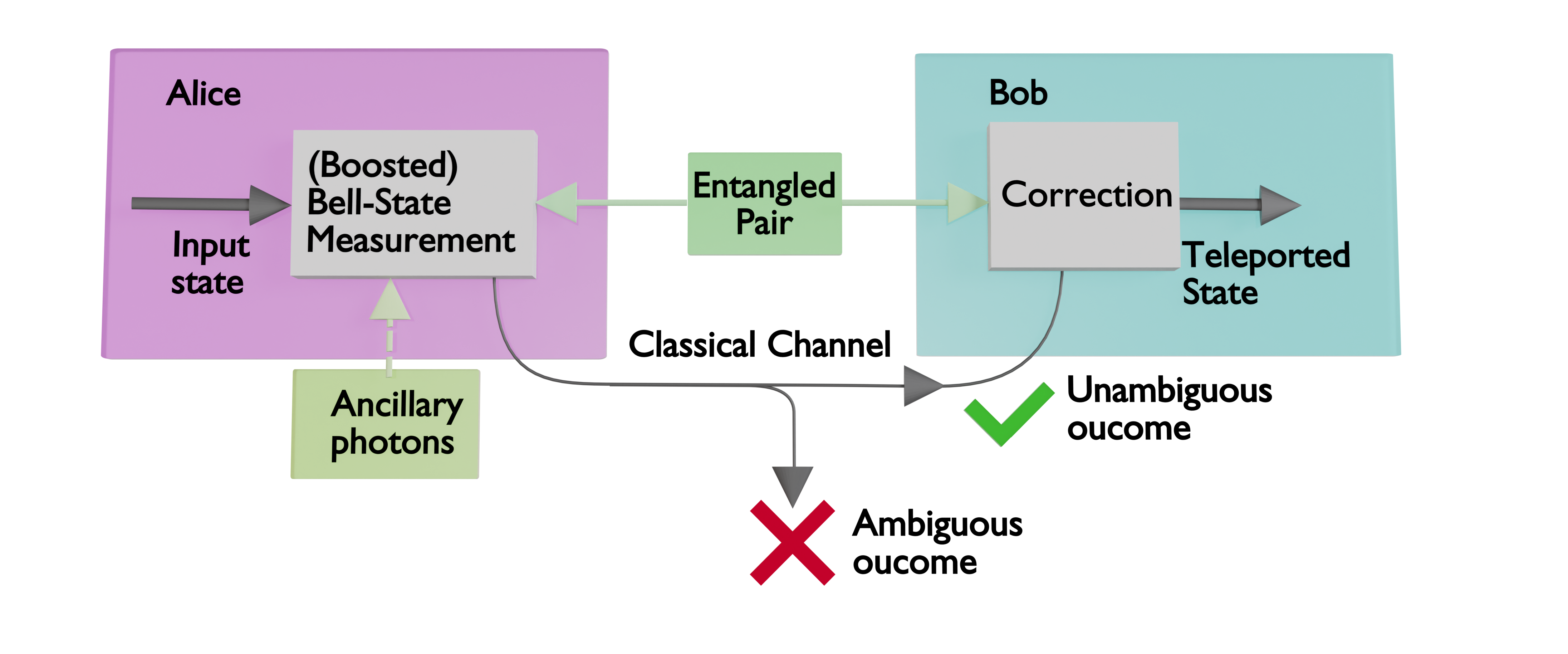}
    \caption{\textbf{Conceptual representation of boosted teleportation protocol.}
    An entangled pair is shared between two parties, Alice and Bob. Alice performs a Bell-State measurement (BSM) between the state to be teleported and one part of the entangled pair, then classically shares the result with Bob. \textbf{a)} The photonic teleportation protocol based on the standard BSM is only successful in $50\%$ of the cases, discarding half of the measurement. \textbf{b)} The use of an additional ancillary state increases the number of unambiguous measurement.}
    \label{fig:Concecpt}
\end{figure*}
In this work, we demonstrate quantum teleportation with a boosted success probability. An ancillary two-photon state is used to increase the probability of a conclusive projective measurement on the Bell-state basis.
We demonstrate that, with current technology, the use of ancillary photons to boost quantum protocols based on teleportation-like operations is a viable option.

\section{Theory}

Quantum teleportation is the process of transferring a quantum state, denoted as $\ket{\Psi_\text{In}}$, from one party (Alice, A) to another party (Bob, B) without the need for a direct transmission of the state. Let us assume Alice holds an input state of the form $\ket{\Psi_\text{In}} = (\alpha \ket{0}_\text{In} + \beta \ket{1}_\text{In})$ in a mode "In". The states $\ket{0}$ and $\ket{1}$ are the logical states, while $\alpha$ and $\beta$ are complex numbers, satisfying $\abs{\alpha}+\abs{\beta}=1$ . The two parties must share entanglement, in the form of a Bell state, defined as:
\begin{equation}
    \begin{aligned}
        \ket{\phi^\pm} &= \frac{1}{\sqrt{2}}\left(\ket{00}\pm\ket{11}\right)\\
        \ket{\psi^\pm} &= \frac{1}{\sqrt{2}}\left(\ket{01}\pm\ket{10}\right)
    \end{aligned}
\end{equation}
In the following, a Bell pair $\ket{\psi^-}$ is shared between the two modes A and B. Together with the state $\ket{\Psi_\text{In}}$, the entire quantum state can then be written as:
\begin{equation}
    \ket{\Psi_\mathrm{Tot}}=\left(\alpha \ket{0}_\text{In} + \beta \ket{1}_\text{In}\right) \otimes \left(\frac{1}{\sqrt{2}}\left(\ket{0_A, 1_B}-\ket{1_A, 0_B}\right)\right) .
\end{equation}
In the first step, Alice performs a joint measurement on the qubits in the modes ``In'' and ``A'', projecting them onto the Bell-state basis. This BSM does not reveal any information about the state $\ket{\Psi_\text{In}}$ itself, but yields two bits of classical information. The state of the qubit B will depend on the measurement outcome and can be written as:
\begin{equation}
    \ket{\Psi_{\mathrm{Out}}}=X^{m_{ZZ}}Z^{m_{XX}}\left(\alpha\ket{0}_B+\beta\ket{1}_B\right) \ 
\end{equation}
where $m_{ZZ},m_{XX} = 0, 1$ denote the outcomes of the BSM.\\
The state $\ket{\Psi_\text{In}}$ can therefore be ``teleported'' onto the qubit B, as long as the outcome of the BSM is known. Alice sends this information to Bob via a classical channel. Bob applies the proper correction and retrieves the teleported state $\ket{\Psi_\text{Tel}}$ in the final step.\\
A quantity that describes the performances of a teleportation is the fidelity $F$ ($0\leq F\leq1$), defined as:
\begin{equation}
    F_\text{T}=\left(\tr \sqrt{\sqrt{\rho_\text{In}}\rho_\text{Tel}\sqrt{\rho_\text{In}}}\right)
\end{equation}
where $\rho_\text{In}$ is the density matrix of the initial and $\rho_\text{Tel}$ the density matrix of the teleported state. This value has an upper limit of $2/3$ for any classical method~\cite{massar1995}. Note that, if the state is an eigenvector of a fixed basis, classical methods can achieve unitary fidelity, so in order to prove a true quantum teleportation, the value of $2/3$ has to be surpassed for any arbitrary vector. An adequate set of states for this purpose is: 
\begin{align}
\begin{split}
    \ket{\psi_\text{In}} \in \bigl\{ &\ket{0},\ \ket{1},\ \ket{+}=\frac{1}{\sqrt{2}}\left( \ket{0}+\ket{1} \right),\\ &\ket{+_i}=\frac{1}{\sqrt{2}} \left(\ket{0} + \mathrm{i} \ket{1}\right) \bigr\}. 
\end{split}
    \label{eq:InputStates}
\end{align}
The teleportation process can be described as a quantum channel, hence it can be represented by a completely positive map $\mathcal{E}$ describing the evolution of the input state. Given an input state $\rho_{In}$ the action of the channel can be written as:
\begin{equation}
\label{eq:process_chi}
    \mathcal{E}(\rho) = \sum_{j,k=0}^{4}\chi_{jk}\hat{A}_j\rho\hat{A}_k^\dagger
\end{equation}
where $\hat{A}_i$ are the Pauli operators and $\chi$ is the matrix describing the process. Studying the action of a channel in the states in equation (Eqn.~\ref{eq:InputStates}) gives information for a complete reconstruction of the process, an operation known as quantum process tomography (QPT)~\cite{obrien2004}. Each BSM outcome $m=(m_{ZZ}, m_{XX})$ can be seen as a different channel, hence it can be associated with a specific $\chi^{m}$. This reconstruction allows for the estimation of the process fidelity as $F_p=\Tr{\chi_{\mathrm{exp}}\chi_{\mathrm{ideal}}}$~\cite{obrien2004}.

\begin{figure*}
    \centering
    \includegraphics[width=1.0\linewidth]{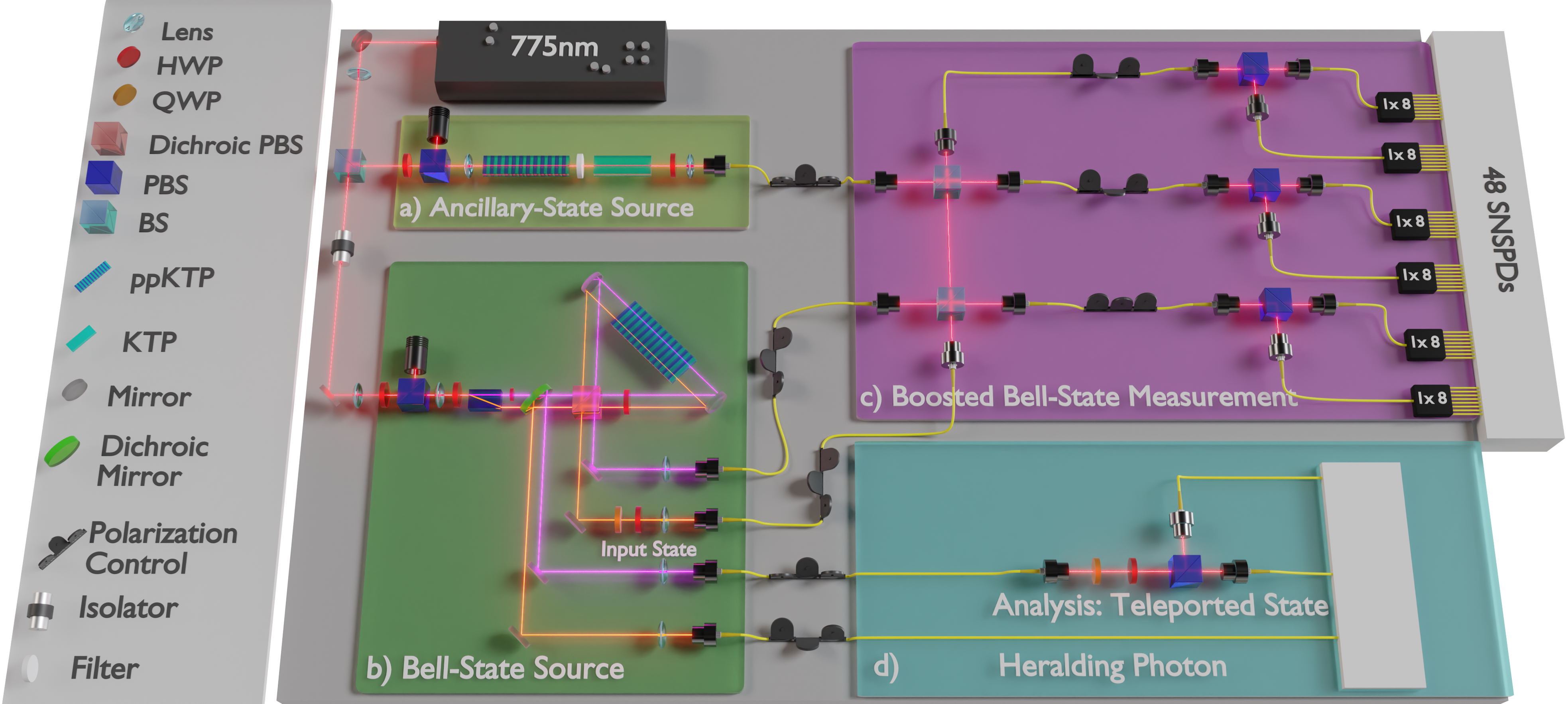}
    \caption{\textbf{Schematic of the experimental set-up.} \textbf{a)} The ancillary state is created via spontaneous parametric down conversion in a nonlinear crystal. The impinging pump beam creates a $\ket{\nu'}=a^\dagger_Ha^\dagger_V\ket{vac}$ photon pair with. A compensation crystal corrects for the temporal walk off and a final half-wave plate (HWP) at $22.5^\circ$ sets the state to $\nu=\frac{1}{\sqrt{2}}({a_H^\dagger}^2-{a_V^\dagger}^2)\ket{vac}$. \textbf{b)} Here, the Bell-state source and the input state source are represented. The different paths are marked as violet and orange, representing the Bell state and the input state, respectively. \textbf{c)} The boosted Bell-State measurement stage projects the input and one qubit from the Bell state into the Bell-state basis. At the second beam splitter these photons interfere with the ancillary state. Translation stages  ensure temporal indistinguishability of the photons. \textbf{d)} The analysis part measures the teleported state. Here, QWPs and HWPs allow for tomographic reconstruction. The lower fiber directs the idler photon of the input sources to a detector to herald the presence of the input state.A detail description can be found in the section experiment set-up.} 
    \label{fig:setup}
\end{figure*}

At the heart of the quantum teleportation protocol is a projective measurement of two photons onto the Bell basis. Here we employ a boosted Bell-state measurement (BBSM) based on a two-photon ancillary state (Fig.\ref{fig:Concecpt}) enabling a success rate of $62.5 \%$~\cite{ewert2014}. Introducing this additional state, more complex output patterns are generated and the success rate is increased. \\
Among the accepted events, i.e. events where the measured click pattern is associated with a Bell State or ambiguous state, we can define the acceptance probability $p_a$ as the ratio of unambiguous measurements over the total amount of events.
If $N_a$ and $N_{amb}$ indicate the respective number of accepted and ambiguous events, the acceptance probability can be estimated as:
\begin{equation}
    p_a = \frac{N_a}{N_a+N_{amb}}.
\end{equation}
For a \textit{full} characterisation of the BSM, a measure of the \textit{quality} of the teleportation is also needed.  Here, we use the fidelity of the teleported state.\\

\begin{figure*}
    \centering
    \includegraphics{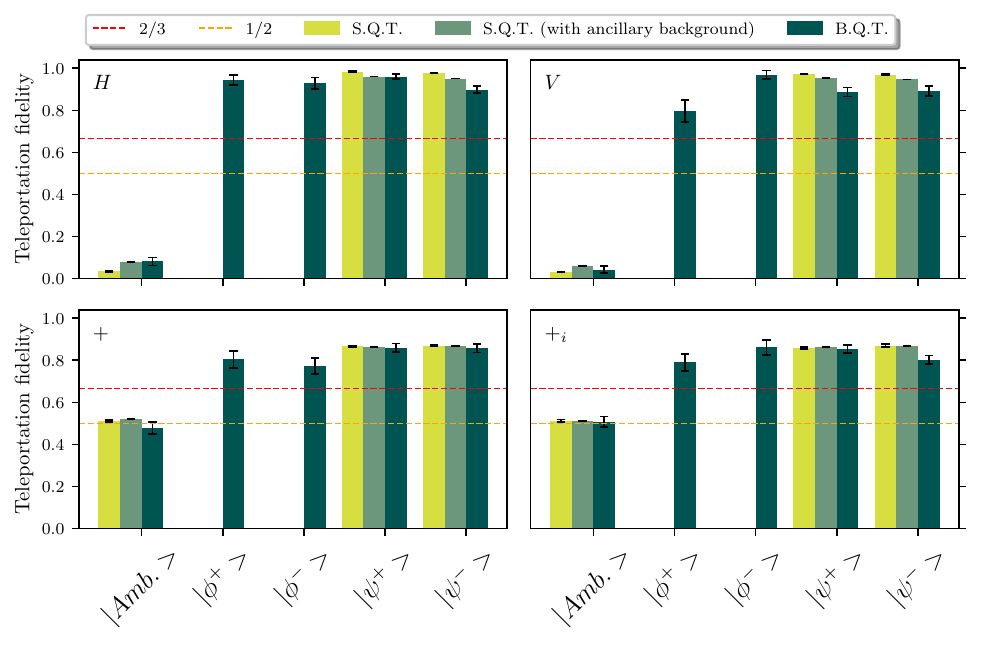}
    \caption{\textbf{Measured fidelities after teleportation.} The figure shows the fidelity between the expected state and the one reconstructed after the teleportation, divided per Bell state detected. Four different input states are tested in three different scenarios: standard teleportation (yellow) and boosted teleportation (dark green). In addition, we extract also the four-photon events corresponding to standard teleportation from the boosted case (light green). The 2/3 limit is shown, that is the best achievable fidelity with classical means, and the 1/2 is also shown as it is what we are expecting in some cases when the BSM is ambiguous. Labels indicate which Bell state was detected.}
    \label{fig:fidelities}
\end{figure*}

\begin{figure}

    \includegraphics[width=0.4\textwidth]{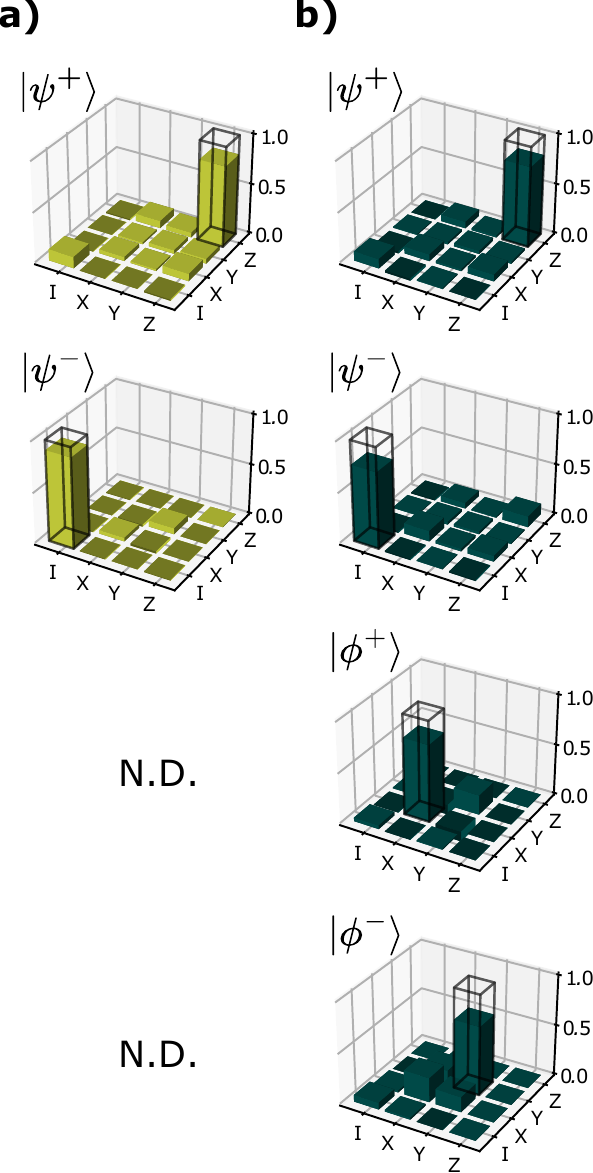}
    \caption{\textbf{Quantum process tomography.} Reconstructed $\chi$ matrices corresponding to each BSM outcome are shown (as defined in Eq.~\ref{eq:process_chi}). Imaginary part is omitted as all terms are close to $0$. Column \textbf{a)} shows the reconstructed process tomographies for the standard BSM, where the states $\phi^+$ and $\phi^-$ are not defined. Column \textbf{b)} shows the results for the boosted teleportation.}
    \label{fig:QPTs}
\end{figure}

\section{Results}
\subsection*{Experimental set-up}
In our experimental implementation, the logical state of the qubit is encoded in the polarisation mode of single photons ($\ket{0}\equiv \ket{H},\ \ket{1}\equiv \ket{V}$).
The photons are generated by parametric down conversion in periodically poled potassium titanyl phosphate crystals (\textbf{ppKTP}) (as shown in Fig. \ref{fig:setup}). A pulsed laser with a wavelength of 775 nm and a pulse duration of 2 ps is used as a pump for the crystal, resulting in the emission of photon pairs degenerated in wavelength at 1550 nm and with orthogonal polarization.\\
To generate the Bell state and input state, one ppKTP crystal is put in a Sagnac-type interferometer (Fig.~\ref{fig:setup} \textbf{b)}).  
 The two sources are created by splitting the pump beam into two parallel paths, which are then fed into the interferometer. The splitting of the path is achieved by having a linearly polarized pump beam at $45^\circ$ that traverses on a beam displacer. The birefringence of the beam displacer causes the two polarisation to separate on a plane parallel to the optical table. Both beams travel through the interferometer parallel to each other, realizing two separate photon sources with only one set of optics. D-shaped half-wave plates allow us to tune the splitting ratio at the PBS independently for the two beams. The Bell-state source is pumped in both the clockwise and counterclockwise direction, generating entangled photon pairs. The second loop on the other hand, is only pumped in the clockwise direction, so that the output state is a heralded single-photon source. A dichroic mirror separates the generated photons from the pump light. Each stage is connected with single-mode fibers along with fiber paddles for polarization compensation.\\
The ancillary state is produced in a linear source (Fig.~\ref{fig:setup} \textbf{a)}), also utilizing a ppKTP crystal. This crystal generates a state $\ket{\nu_0}=a^\dagger_Ha^\dagger_V\ket{vac}$ which, by means of a half-wave plate at $22.5^\circ$, becomes $\nu=\frac{1}{\sqrt{2}}({a_H^\dagger}^2-{a_V^\dagger}^2)\ket{vac}$~\cite{ewert2014}.\\
The BBSM is realized by two balanced non-polarizing beam splitters (Fig.~\ref{fig:setup} \textbf{c)}). The first combines the input state with one of the Bell-state photons, while the second one overlaps one of the outgoing modes of the first beam splitter and the ancillary state. The outputs of the interferometer are filtered by in-line fiber filters with a width of $1.5\ \text{nm}$ to eliminate any unwanted photons from the pump laser or environment. Photons of different polarization in each mode are separated with a polarising beam splitter. The two modes are then evenly divided into eight modes with a fiber based one-by-eight splitter. Each output is directed onto a superconducting nano-wire single-photon detector. This allows a pseudo-photon number resolving measurement.\\
The second photon from the Bell state is sent directly to a tomography stage (Fig.~\ref{fig:setup} \textbf{d)}) to analyze the properties of the teleported state. Here, wave plates, a polarizing beam splitter, and two detectors allow us to project the photon's state in any Pauli basis and perform a full state tomography on the teleported state.\\
The data collected from the detectors is analysed, and events where a herald and a second Bell-state photon are present are post-selected. From this data set, only events with a total photon number of four or six photons are further processed, depending on the protocol. 
For the analysis of the non-boosted protocol events with four-photons, comprising one photon pair each from both the Bell-state source and input-state source, are used. The evaluation of boosted teleportation is done with six-fold coincidences, consisting of one Bell-state pair, one input-state pair, and an ancillary state.

\subsection*{Measurement results}
Before studying the teleportation process, the performances of the three sources are evaluated independently. First, the quality of the generated input states (Eqs.~\ref{eq:InputStates}) is assessed, performing state tomographies. The reconstructed density matrices show that the fidelities of the single-qubit is on average \textcolor{black}{$\Bar{F} = 0.9958\pm0.0017$}. The Bell state used in the teleportation is characterized in a similar way. Quantum state tomography is performed on the Bell state by measuring correlations between the two photons in different polarization bases. The density matrix of the state indicates a state fidelity of $97.45(2)\%$, the reduction being caused by higher-order emissions and experimental imperfections.
The estimation of the density matrix is done by mean of maximum likelihood estimation and the provided uncertainty is calculated assuming Poissonian statistics.\\
For the characterization of the ancillary state, standard tomography methods cannot be applied, as the two photons are in the same spatial mode. For characterising the state, we rotate an half-wave plate to continuously change from $\nu_0=a_H^\dagger a_V^\dagger \ket{vac}$ to $\nu=\frac{1}{\sqrt{2}}\left({a_H^\dagger}^2-{a_V^\dagger}^2\right)\ket{vac}$. Monitoring coincidences ($CC$) between $H$ and $V$ as a function of the wave plates angle, we can define the visibility as:
\begin{equation}
    V_{\ket{\nu}} = \frac{\max{CC_{HV}}-\min{CC_{HV}}}{\max{CC_{HV}}}, 
\end{equation}
which gives an indication for the quality of the state. In our measurements this visibility shows a value of $V_{\ket{\nu}}=98.01(13)\%$.\\
After initial characterization, teleportation is performed. For each of the input states, the teleported qubit on Bob's side is measured in the bases X, Y, and Z. All possible coincidence patterns between the detectors are recorded, and counts of all click patterns up to a total number of six photons are stored.
Each click pattern is indicative for one of the Bell states ($\ket{\Psi^\pm},\ \ket{\Phi^\pm}$) or an ambiguous outcome. We sort all measurement events by these states and get five different sets of data for each teleportation. A maximum likelihood estimation is performed on each set, obtaining an estimation of the corresponding teleported density matrix. Once data for all four input states is collected, QPT is performed to reconstruct the quantum channel $\chi$ matrices.\\
We study the behaviour of the teleportation fidelity in three different scenarios, depending on the presence or absence of the ancillary state.

\paragraph{Standard quantum teleportation.} The first method is the standard quantum teleportation (SQT), obtained by blocking the ancillary state source and postselecting for four-photon events. This measurement is an unboosted BSM with a maximal efficiency of $50\%$, with the only difference that an additional beam splitter is in one of the paths~\cite{Braunstein1995}. Note that the second beam splitter does not affect the measurement. The resulting fidelities of the teleported states are shown in Fig.~\ref{fig:fidelities} (yellow bars). \\
The teleported state shows an average fidelity of \textcolor{black}{$F^{ST}_T = 0.9215\pm0.0012$} for $\ket{\Psi^\pm}$. When the outcome is ambiguous, however, it is not defined which correction should be applied. In fact it is not known if the outcome of the BSM would be the state $\ket{\Phi^+}$ or $\ket{\Phi^-}$, meaning that with a $50\%$ chance the correct gate is either $\hat{X}$ or $\hat{Y}$. This effect manifests differently for different input states. While for $H$ and $V$ the fidelity of the ambiguous case is \textcolor{black}{$F_T^H = 0.0334\pm0.002$} and \textcolor{black}{$F_T^V = 0.00306\pm0.002$} respectively (as for both gates the teleported state is orthogonal to the input state), for $\ket{+}$ and $\ket{+_i}$ this is respectively \textcolor{black}{$F^+_T=0.510\pm0.004$} and \textcolor{black}{$F^{+_i}_T=0.512\pm0.006$}. The reconstructed process for the unambiguous results shows a process fidelity of \textcolor{black}{$F_p=0.8613\pm0.0047$}.
An estimation of the acceptance probability results in a \textcolor{black}{$p_a=52.74 \pm 0.19\%$}, in line with the theory bound. By multiplying this value with the average state fidelity, we can obtain a measure of the quality $q = p_a \cdot F_T$ for the teleportation. Here, we achieve a value of \textcolor{black}{$q_{S.T.}=48.60 \pm 0.18 \%$}.\\

\begin{figure}
    \centering
    \includegraphics[width=1\linewidth]{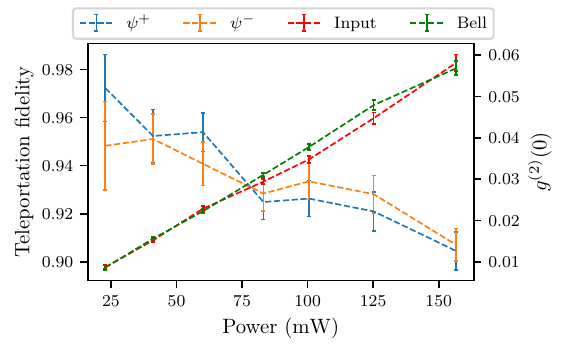}
    \caption{\textbf{Fidelity reduction due to non-perfect interference.} In the plot fidelities and $g^{2}(0)$ as a function of power are shown. The teleported state is a $\ket{+}$ state. \textbf{Green and Red lines. }As a function of the power, the measured $g^{2}(0)$ value increases, meaning more higher-order emissions from the sources. \textbf{Blue and orange lines.} Fidelity of the teleported state in the standard scenario is measured for different total power of the Bell and Input sources.Higher-orders emissions result in worse visibility of BSM, reducing the overall fidelity of the teleportation. Only the results for $\ket{\psi^\pm}$ are shown as the ambiguous measurement are not influenced by this effect and the $\ket{\phi^\pm}$ states are not defined.
    }
    \label{fig:effect_of_visibility}
\end{figure}

\paragraph{Boosted quantum teleportation.} Next, a boosted quantum teleportation (BQT) is implemented, by allowing the ancillary state into the setup. The teleported fidelities are represented by dark green bars in Fig.~\ref{fig:fidelities}. Each input state is measured with a total integration time of roughly $50h$. The fidelities of the teleported state are all above the classical limit of $2/3$, proving true quantum teleportation also for the $\ket{\Phi^\pm}$ BSM outcomes. Averaging the fidelities over all the unambiguous cases, we obtain a value of \textcolor{black}{$F_T^{BQT}=0.8677\pm0.0024$}. The process fidelity in this case is \textcolor{black}{$F_p^{BQT}=0.773\pm0.016$}.
The acceptance probability of the Bell-state measurement increases from \textcolor{black}{$p_a=52.74 \pm 0.19\%$} of the S.T. to \textcolor{black}{$p_a=69.71\pm 0.75\%$}. It exceeds the theoretical limit of $62.5\%$ because incorrectly identified states are still counted as successful measurements, even though the outcome is not correct. Again, we calculate the product of the acceptance probability and the average fidelity. For the boosted teleportation we achieve a value of $q_{B.T.}=60.48 \pm 0.67 \%$, while the theoretical limit is $0.625$. This shows improvement of the boosted teleportation compared to the standard teleportation.\\
\paragraph{Standard quantum teleportation with background.} For completeness, a standard teleportation is extracted from the four-photon event of the BQT scenario. If ancillary-state generation fails, a normal unboosted teleportation will take place. The set-up then performs a standard BSM, analogous to the one used in the SQT scenario. However, in this case, the emission of the ancillary state source is still present, causing background events that, together with photon loss, can result in some events being indistinguishable from the desired measurement. The light green bars in Fig.~\ref{fig:fidelities} are displaying the measured fidelities for this case. The average fidelity is \textcolor{black}{$F_T^{SQT_b}=0.90944\pm0.00006$} the average of the estimated process fidelities is \textcolor{black}{$F_p^{SQT_b}=0.85653\pm0.00078$}.
Fig. \ref{fig:QPTs} shows the reconstructed $\chi$ matrices for the SQT and BQT divided per BSM outcome.\newline
The fidelities resulting from this measurement are above the classical threshold. The difference between the SQT and BQT in fidelity is mostly caused by multi-photon emission combined with losses in the set-up. The used SPDC source are also emitting unwanted higher photon numbers. These states will interfere with the BSM and cause incorrect measurement outcomes. To show the influence of this phenomenon, a standard teleportation, using the input state $\ket{+}$, is performed at different pump powers. The results are shown in Fig.~\ref{fig:effect_of_visibility}. The fidelity of the state $\ket{+}$ is plotted as a function of the pump power for $\ket{\psi^+}$ and $\ket{\psi^-}$ (orange and blue bar). A measure of the higher order content of the produced squeezed state is the second order correlation function $g^2(0)$. Its measured value is also represented in green and red, respectively for the Bell state and input sources. This indicates that the main source of noise originates are higher-order emissions of the SPDC sources.

\section{Discussion}

In this work, we demonstrate a boosted quantum teleportation protocol.
Incorporating an additional ancillary two-photons state, we achieve an improved acceptance rate of \textcolor{black}{$p_a=69.71\pm 0.75\%$} (compared to \textcolor{black}{$p_a=52.74 \pm 0.19\%$} in the standard scenario). This surpasses the theoretical limit of $50\%$ for linear-optical set-ups. While the use of more photons introduces more experimental imperfections, our results shows that this only has a minimal effect on the fidelity of the teleported state. The mean state fidelity after teleportation is \textcolor{black}{$F_T^{BQT}=0.8677\pm0.0024$}, well above the threshold of $2/3$. To the best of our knowledge, this is the first linear-optical teleportation with an efficiency of more than $50\%$.
Efficient quantum teleportation is crucial for the development of future quantum networks. The set-up operates at the telecommunication C-band ($1550\ $nm), which makes it suitable to integrate these technologies into existing optical networks. Due to the use of only linear-optical elements, the scheme can be scaled in a straightforward manner. It can be generalized to achieve efficiency arbitrarily close to $100\%$ by using additional beam splitter and ancillary states. 
Our findings are relevant in the context of quantum repeaters, where even minor enhancements in success rates can result in exponential improvements for large repeater chains. In addition, measurement-based quantum computers have the potential to benefit from this scheme, as it reduces the loss of information caused by failed teleportations.

\begin{footnotesize}
\noindent
\\
\noindent\textbf{Acknowledgements}
\textcolor{black}{We thank Nico Hauser and Lukas Rückle for discussions and comments. We would also like to thank Jakob Budde for his work on the draft of Fig. \ref{fig:setup} and Shreya Kumar for her input on the graphs}.
We would like to acknowledge support from the Carl Zeiss Foundation, the Centre for Integrated Quantum Science and Technology (IQST), the Deutsche Forschungsgemeinschaft (DFG, German Research Foundation) - 431314977/GRK2642, the Federal Ministry of Education and Research (BMBF, project SiSiQ and PhotonQ).
\end{footnotesize}

\newpage
\clearpage
\onecolumngrid
\appendix

\section{Process fidelities values}

\begin{table}[h]
    \centering
   \begin{tabular}{c|c|c|c}
         BSM outcome&  SQT&  SQT with background& BQT\\ \hline
         $\ket{\psi^+}$&  0.8082(46)&  0.8113(15)& 0.807(20)\\
         $\ket{\psi^-}$&  0.9144(82)&  0.90178(4)& 0.770(20)\\
         $\ket{\phi^+}$&  N.D.&  N.D& 0.735(43)\\
         $\ket{\phi^-}$&  N.D.&  N.D.& 0.781(37)\\
    \end{tabular}
    \caption{Reconstructed process fidelities}
    \label{tab:my_label}
\end{table}

\end{document}